\documentclass[9pt,twocolumn,twoside]{opticajnl}
\journal{opticajournal} 

\setboolean{shortarticle}{true}

\usepackage{lineno}

\title{Resonantly enhanced terahertz four-wave mixing in fluorides}

\author[1,2]{Eva Noskovicova}
\author[1]{Martin Koys}
\author[1,2]{Monika Jerigova}
\author[1]{Dusan Velic}
\author[1,*]{Dusan Lorenc}

\affil[1]{International Laser Centre, Ilkovicova 3, 84104 Bratislava, Slovakia}
\affil[2]{Department of Physical and Computational Chemistry, Comenius University, Ilkovicova 6, Bratislava, Slovakakia}

\affil[*]{dusan.lorenc@gmail.com}

\begin{abstract}

Converting a THz signal into the optical domain is of great interest for THz sensing and spectroscopy. Here intense broadband THz pulses with a central frequency of $\Omega$$_{THz}$ are mixed with an optical pump at $\omega$$_p$ and a signal is observed at a wavelength of $\omega$$_s$ = 2($\omega$$_p$ - $\Delta\omega$$_p$) - $\Omega_{THz}$ with the detuning $\omega$$_p$ - $\Delta\omega$$_p$ being due to pump pulse spectral broadening. The observed THz-FWM signal close to 400 nm is shown to result from a resonantly amplified four-wave mixing in CaF$_2$, BaF$_2$ and MgF$_2$. 

\end{abstract}

\setboolean{displaycopyright}{false} 

\begin{document}

\maketitle

Rapid and sensitive THz-frequency gas- and molecule- detectors – typically operating in the range of 0.1-10 THz – are at the forefront of current research effort as they would ideally allow an unambiguous identification of chemicals and compounds at extremely low concentration levels – potentially down to the single molecule scale. THz based gas detection, for example, shows great promise toward both environmental and security related applications \cite{2014}. The THz frequency domain is ideally suited for sensing small molecules and larger hydrocarbons, since they typically show strong response within this spectral range \cite{PICKETT1998}. However, current THz spectrometers are typically cumbersome and costly, since they either rely on mechanical delay stages, or coupled/synchronized femtosecond laser sources. Hence converting the actual THz signal into the optical domain would be of great merit. Formerly, nonlinear wave-mixing has been extensively explored especially in the gas phase \cite{Cook2000, Zhang2010} but reports on terahertz four-wave mixing (FWM) in solid state were rather scarce \cite{Suizu2007, Wu2011, Clerici2013, Koys2017}. In the very recent paper by Le et. al. \cite{Le2023} a THz generation was demonstrated through Raman enhanced FWM in diamond.  The approach while being of great interest requires two optical parametric amplifiers along with stretchers and hence a rather complicated setup.
Here we explore THz-FWM in fluorides as a mean of converting the THz signal into the optical domain. Experimentally our approach follows that of Clerici et. al. \cite{Clerici2013} but instead of diamond, we explore fluorides (CaF$_2$, MgF$_2$ and BaF$_2$) and show resonant enhancement of the FWM. The approach is simple and experimentally straightforward to implement without the requirement of multiple laser sources.

\begin{figure}[ht!]
\centering
\includegraphics[width=\linewidth]{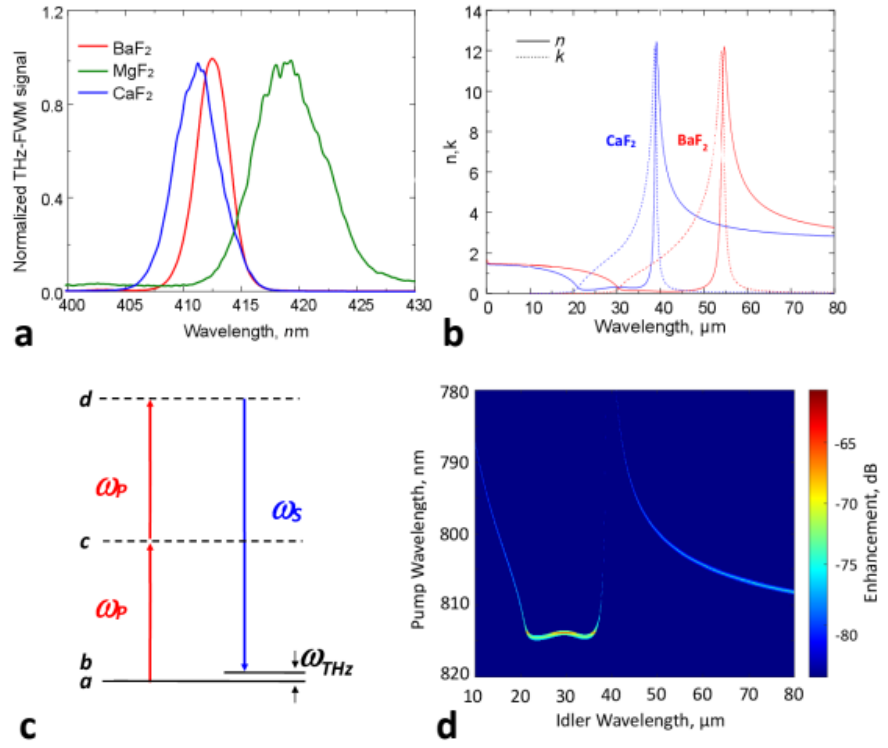}
\caption{THz-FWM signal spectra as obtained in CaF$_2$, BaF$_2$ and MgF$_2$ a). Material dispersion and absorption index of CaF$_2$ and BaF$_2$ in the THz range extracted from ref. \cite{Kaiser1962} b). Schematic depiction of the THz-FWM process c). Calculated phase-matching condition for THz-FWM in CaF$_2$ d).}
\label{fig:init}
\end{figure}

The output of an amplified Ti:Sapphire system (COHERENT Legend DUO) was split into two beams with the main output of 3 mJ at 120 fs serving as a pump for broadband 2-colour plasma based THz source [4] and the remaining 0.5 mJ used as an optical pump for the FWM. Both the pump and THz idler pulses were overlapped inside of the fluoride sample and a combination of high-pass filter (Standa UAB) and a fiber coupled spectrometer (Flame-T, Ocean Optics) was used to sample the THz-FWM optical signal. A set of optical grade microscopic cover slips was used to tune the pump power and a combination of two wire-grid polarizers (Pure-Wave Polarizers) was employed as a variable attenuator for the THz idler beam.  
By combining the 800 nm pump beam with the THz idler in the forward (co-propagating) geometry, a THz-FWM signal has been observed in the optical range as shown in Fig~\ref{fig:init}a. Note that we’ve observed THz-FWM in all three available fluoride samples i.e. CaF$_2$, BaF$_2$ and MgF$_2$ showing signal peaks at different wavelengths and hinting at material dependent phase-matching mechanism, see the discussion below. For the sake of consistency, other samples were tested as well including fused silica, CVD diamond, sapphire, LiNbO$_3$, THz compatible plastics (Zeonex, Zeonor, Topas) but neither provided an observable effect. 

Given the centrosymmetric nature of the fluoride samples we can generally exclude contribution of even-order processes to the observed signal and the fact, that the signal was only observable in selected fluorides would hint at material specific enhanced $\chi$$^{(3)}$.

Let’s now turn our attention to the spectral order of THz-FWM signals observed in respective fluorides as was shown in Fig~\ref{fig:init}a. The seminal work of Kaiser et. al. \cite{Kaiser1962} on far infrared optical properties of BaF$_2$, CaF$_2$ and SrF$_2$ had indicated strong resonances in the 5.5 – 8 THz range that were ascribed to transverse optical (TO) phonons of the host lattice as depicted in Fig~\ref{fig:init}b. A recent work \cite{Kaplunov2021} confirmed the former results and more importantly extended the far infrared/THz absorption data to MgF$_2$. Hence a resonance enhanced FWM in fluorides should be observed given the available THz source can cover the required range of frequencies. In the following we will focus on BaF$_2$, CaF$_2$ as to our knowledge there is currently no reliable data on refractive index of MgF$_2$ in the far infrared/THz range.

For an efficient FWM to occur, phase matching condition has to be satisfied:

\begin{equation}
2\mathbf{\mathit{k}}_{p} = \mathbf{\mathit{k}}_{s} + \mathbf{\mathit{k}}_{THz}
\label{eq:refname1}
\end{equation}

where \textit{k}$_p$, \textit{k}$_s$ and \textit{k}$_{THz}$ are respective wave vectors of pump, signal and the (THz) idler waves as schematically depicted in Fig~\ref{fig:init}c. This condition is generally difficult to attain given the broad range of frequencies involved (10$^{12}$-10$^{15}$ Hz) although some benefits can be gained in the counter-propagating geometry \cite{Clerici2013}. Previously \cite{Andrews1981}, it had been shown that in the presence of strong resonances the expected signal intensity is given as:

\begin{equation}
\mathit{I}_{FWM}=(\mathit{L}/2)^{2}exp(-\alpha\mathit{L} )\left | \chi^{(3)}  \right |^{2}G(\Delta k)
\label{eq:refname2}
\end{equation}

where:

\begin{equation}
G(\Delta k) = \frac{1+exp(-\Delta\alpha\mathit{L})-2exp(-{\Delta}\alpha\mathit{L}/2)cos({\Delta} kL)}{({\Delta} k^{2}+{\Delta}\alpha ^{2}/4)(L/2)^{2}}
\label{eq:refname3}
\end{equation}

\begin{equation}
\Delta k = 2{\mathit{k}}_{p} - {\mathit{k}}_{s} - {\mathit{k}}_{THz}
\end{equation}

\begin{equation}
\Delta \alpha = 2{\alpha_{p}} - {\alpha_{s}} + {\alpha_{THz}}
\end{equation}

with L being the sample thickness, and  $\alpha$ being the respective absorption coefficients. Here we have dropped the vector nature of wave vectors as we are only interested into the collinear interaction. Note that the $\chi$$^{(3)}$ itself is a resonantly enhanced quantity, hence: 

\begin{equation}
{\chi}_{ijkl}^{(3)} = \frac{A_{ijkl}}{(\omega_{ba}-(2\omega_p-\omega_s)-i\Gamma _{ba})} + {\chi}_{NR}^{(3)}
\end{equation}

with $\chi$$^{(3)}$$_{NR}$ being the nonresonant part of $\chi$$^{(3)}$. Making use of (1) and ignoring the nonresonant part we calculated the FWM intensity I$_{FWM}$ for CaF$_2$ as shown in Fig~\ref{fig:init}d. and BaF$_2$ (supplementary information)  as a function of $\lambda$$_p$ and $\lambda$$_{THz}$ with the resonant enhancement given by the color bar in the relative units. Note that $\lambda$$_s$ was fixed to the experimentally observed values as shown in Fig~\ref{fig:init}a. Consequently for CaF$_2$ by using a pump wavelength in the vicinity of 814 nm an enhanced FWM signal is expected for the THz idler in the 22-38 $\mu$m (8-14 THz) range. As shown in the supplementary for BaF$_2$, the PM conditions imply a pump wavelength close to 814 nm and the THz idler in the 30-50 $\mu$m (6-10 THz) range. One clear conclusion that can be directly drawn from PM conditions is that a broadband THz source is required spanning the wavelengths/frequencies required for PM. It has been previously shown \cite{Koulouklidis2016} that the 2-color (air) plasma THz sources with their bandwidth exceeding 20 THz are matching the requirements. Note that in our experiment we have used a dedicated THz low-pass filter to cut-off wavelentghts below approximately 20 $\mu$m from participating in the THz-FWM process (see supplementary material for more details).   

\begin{figure}[ht!]
\centering
\includegraphics[width=\linewidth]{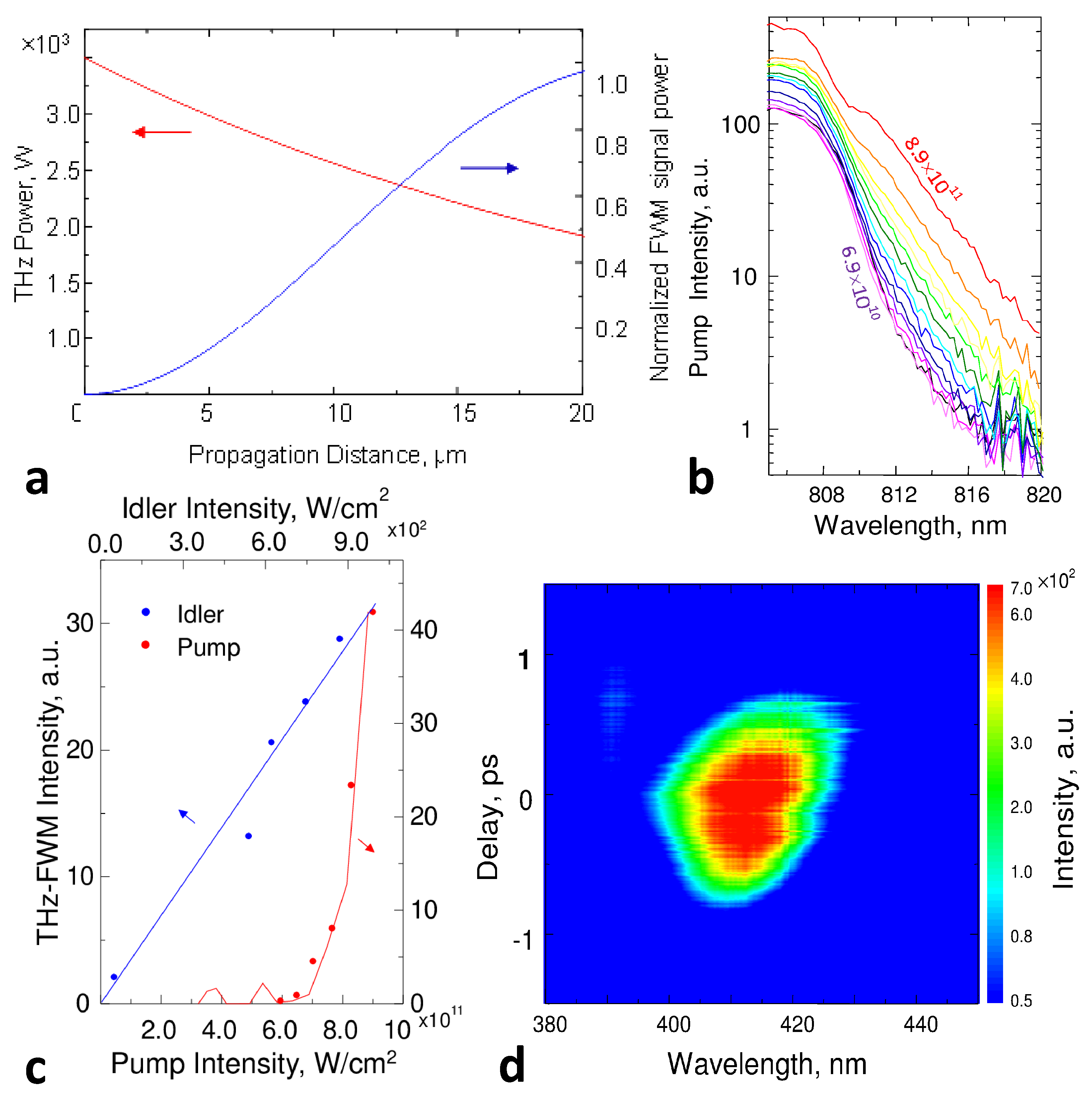}
\caption{THz-FWM signal power and the THZ idler power as a function of the propagation distance inside of the CaF$_2$ sample a). Pump spectral broadening as a function of pump intensity (the limits are shown in units of W/cm$^2$) b). THz-FWM signal as a function of pump and idler intensity c). Spectrogram of the THz-FWM signal d).}
\label{fig:mid}
\end{figure}

Now that we have identified the PM wavelengths with respect to pump and idler, we can return to Fig~\ref{fig:init}b. The terahertz idler wavelength is located in the near-zero index (NZI) \cite{Kinsey2019} spectral region with all the consequences for idler propagating in the slow light regime. However both pump and signal being out of NZI region do experience appreciable finite dispersion and group indices resulting into a walk-off distance of approximately 20 $\mu$m between the optical fields and the terahertz field thereby limiting the useful interaction length. On the other hand, the THz also experiences large absorption losses as is evident from the dashed lines in Fig~\ref{fig:init}b. Hence we anticipate the FWM interaction to take place within a couple of dozen micrometers in the high-loss anomalous dispersion part of the THz spectrum.  

Previous conclusions have been verified and confirmed by solving a set of coupled generalized nonlinear Schroedinger equations (CGNLSE). The approach is based upon our own earlier work \cite{Koys2017} and allows to couple optical and THz fields using a broadband solver. Fig~\ref{fig:mid}a shows the THz seed power being depleted as it propagates through the sample while at the same time there is an observable onset of the THz-FWM signal. Note that the signal reaches a maximum at around 20 $\mu$m from the edge of the sample in line with the former qualitative arguments.

To substantiate these findings we’ve experimentally checked for the intensity scaling of the THz-FWM signal with respect to the pump and THz idler intensity in CaF$_2$. Naturally, a  FWM process as depicted in Fig~\ref{fig:init}a is expected to follow the scaling: $I_{sig}$ = $I_{THz}$$I_{pump}$$^2$.  As shown in Fig~\ref{fig:mid}c the FWM signal clearly follows linear dependence upon the THz idler intensity however the scaling with respect to pump intensity is far from being parabolic. Instead it shows activated behavior with a threshold around 6×10$^{10}$ W/cm$^2$ and a rapid rise thereafter. As will be shown below this is to be traced back to the combined effect of pump spectral broadening and FWM.

As the pump pulse with a central wavelength of 800nm propagates through the medium in the normal dispersion regime it undergoes spectral broadening mostly due to an interplay between self-phase modulation and group velocity dispersion \cite{Agrawal}. Hence a nonzero tail appears in the sub 820nm range of the pump spectrum as depicted in Fig~\ref{fig:mid}b. With a further increase of the intensity this will subsequently act as new frequency-shifted pump.  After we have adopted the 814nm resulting from the phase-matching analysis as a new pump source and applied the assumed scaling of  $I_{sig}$ = $I_{THz}$$I_{pump}$$^2$ an excellent match was achieved with the experimental data as shown by solid lines in Fig~\ref{fig:mid}b.
Finally, we’d like to draw a direct comparison to the works of Clerici et. al. and Le et. al. \cite{Clerici2013, Le2023}. To account for the former a time resolved spectrogram was constructed by delaying the pump vs. idler and registering signal spectra at each respective step. The data in Fig~\ref{fig:mid}d show a well localized temporal pulse with a pulse duration of 500 fs FWHM thereby excluding long lasting coherent effects. This is also in stark contrast to the previous study on diamond \cite{Clerici2013} where an 8 ps long lasting component has been observed and it was interpreted as efficient backward phase matched interaction seeded by the terahertz signal. Hence we can safely exclude effects of backward propagating fields in the current study. 
As for the latter \cite{Le2023} because of the non-collinear phase-matching the authors have observed emission of different THz components in different directions with different angular spreads. However the signal beam generated in fluorides is directional and coaxial with the pump beam with a measured beam divergence of 5 mrad limited by the pump beam divergence. The far-field beam profile of the signal beam as imaged by a camera is shown in the supplementary. It shows a radially symmetric low-order mode profile. Consequently we have excluded non-collinear phase matching \cite{Le2023} having any significant contribution to the signal.

The observed resonant enhancement of THz-FWM in fluorides opens up new possibilities for the THz/optical conversion in a range of materials showing resonances in the THz range and given favourable PM conditions are met. In the former work of Le et.al. \cite{Le2023} diamond was suggested as a material of choice for FWM based THz generation but our work is a direct proof that other options should not be discarded.      

We have demonstrated THz-FWM in selected fluorides BaF$_2$, CaF$_2$, MgF$_2$ and identified the effect to be due to resonant enhancement. The phase matching was shown to occur between a frequency shifted pump and a THz idler located in the vicinity of a strong resonance. Our approach shows potential toward THz detection by optical means.

\begin{backmatter}
\bmsection{Funding} 
This research has been supported by NATO SPS program under project G5795.

\bmsection{Acknowledgments} We are grateful for fruitful discussions with Simon Fleming and Alessandro Tuniz.

\bmsection{Disclosures} The authors declare no conflicts of interest.

\bmsection{Data availability} Data underlying the results presented in this paper are not publicly available at this time but may be obtained from the authors upon reasonable request.


\bmsection{Supplemental document}
See Supplement 1 for supporting content. 

\end{backmatter}

\bibliography{sample}

\begin{thebibliography}{10}
\newcommand{\enquote}[1]{``#1''}

\bibitem{2014}
\emph{Terahertz and Mid Infrared Radiation: Detection of Explosives and CBRN (Using Terahertz)} (Springer Netherlands, 2014).

\bibitem{PICKETT1998}
H.~PICKETT, R.~POYNTER, E.~COHEN, \emph{et~al.}, {\protect\JournalTitle{Journal of Quantitative Spectroscopy and Radiative Transfer}} \textbf{60}, 883–890 (1998).

\bibitem{Cook2000}
D.~J. Cook and R.~M. Hochstrasser, {\protect\JournalTitle{Optics Letters}} \textbf{25}, 1210 (2000).

\bibitem{Zhang2010}
X.-C. Zhang and J.~Xu, \emph{Introduction to THz Wave Photonics} (Springer US, 2010).

\bibitem{Suizu2007}
K.~Suizu and K.~Kawase, {\protect\JournalTitle{Optics Letters}} \textbf{32}, 2990 (2007).

\bibitem{Wu2011}
H.~Wu, H.~Liu, N.~Huang, \emph{et~al.}, {\protect\JournalTitle{Applied Optics}} \textbf{50}, 5338 (2011).

\bibitem{Clerici2013}
M.~Clerici, L.~Caspani, E.~Rubino, \emph{et~al.}, {\protect\JournalTitle{Optics Letters}} \textbf{38}, 178 (2013).

\bibitem{Koys2017}
M.~Koys, E.~Noskovicova, D.~Velic, and D.~Lorenc, {\protect\JournalTitle{Optics Express}} \textbf{25}, 13872 (2017).

\bibitem{Le2023}
J.~Le, Y.~Su, C.~Tian, \emph{et~al.}, {\protect\JournalTitle{Light: Science \&amp; Applications}} \textbf{12} (2023).

\bibitem{Kaiser1962}
W.~Kaiser, W.~G. Spitzer, R.~H. Kaiser, and L.~E. Howarth, {\protect\JournalTitle{Physical Review}} \textbf{127}, 1950–1954 (1962).

\bibitem{Kaplunov2021}
I.~A. Kaplunov, G.~I. Kropotov, V.~E. Rogalin, and A.~A. Shakhmin, {\protect\JournalTitle{Optical Materials}} \textbf{115}, 111019 (2021).

\bibitem{Andrews1981}
J.~Andrews, R.~Hochstrasser, and H.~Trommsdorff, {\protect\JournalTitle{Chemical Physics}} \textbf{62}, 87–101 (1981).

\bibitem{Koulouklidis2016}
A.~D. Koulouklidis, V.~Y. Fedorov, and S.~Tzortzakis, {\protect\JournalTitle{Physical Review A}} \textbf{93} (2016).

\bibitem{Kinsey2019}
N.~Kinsey, C.~DeVault, A.~Boltasseva, and V.~M. Shalaev, {\protect\JournalTitle{Nature Reviews Materials}} \textbf{4}, 742–760 (2019).

\bibitem{Agrawal}
G.~P. Agrawal, \emph{Nonlinear Fiber Optics} (Springer Berlin Heidelberg, 2019).

\end{thebibliography}

\bigskip
\pagebreak
\bigskip

\section{Resonantly enhanced terahertz four-wave mixing in fluorides. SUPPLEMENTARY INFORMATION}

\maketitle


\subsection{Phase-Matching in BaF$_2$}

\begin{figure}[ht!]
\centering
\includegraphics[width=\linewidth]{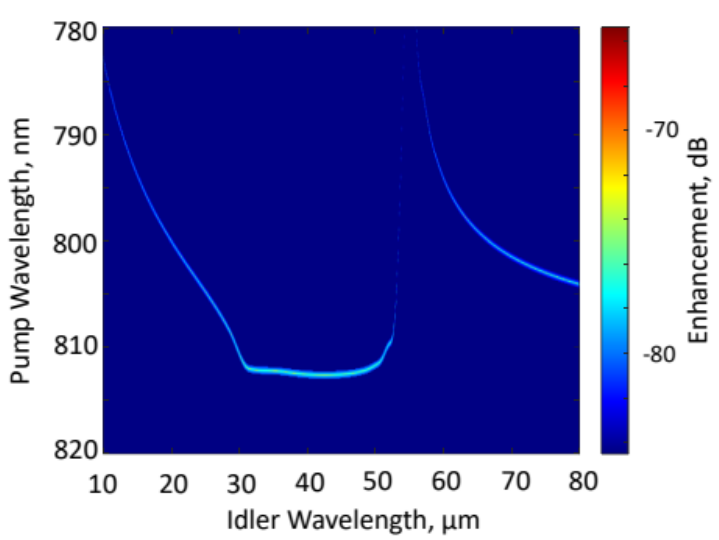}
\caption{Calculated phase-matching condition for THz-FWM in BaF$_2$.}
\label{fig:PM}
\end{figure}

Using the formalism described in the main text and following Eq(1-6) we calculated the PM conditions for BaF$_2$ as depicted in Fig. S~\ref{fig:PM}. The PM conditions imply a pump wavelength close to 814 nm and the THz idler in the 30-50 $\mu$m (6-10 THz) range. This is, as expected, qualitatively similar to the case of CaF$_2$ with the idler (THz) red shifted as compared to the latter.

\subsection{THz-FWM mode profile}

The THz-FWM profile has been captured by means of an SLR camera (Canon EOS6D MarkII) at the distance of 400mm from the sample surface. It shows a low-order transverse mode with a calculated beam divergence of 5 mrad.

\begin{figure}[!ht]
\centering
\includegraphics[width=0.7\linewidth]{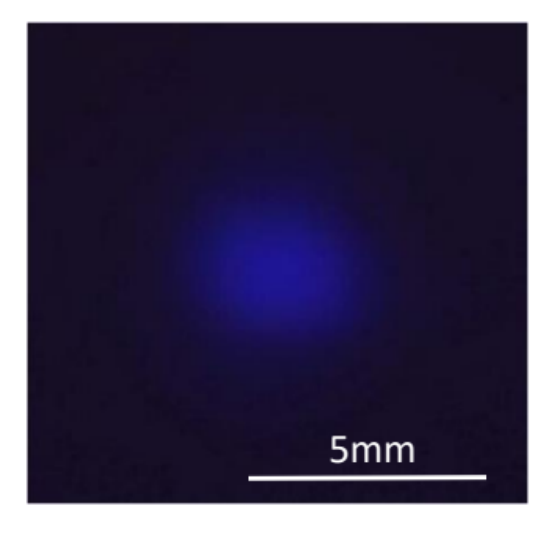}
\caption{Far field THz-FWM mode profile captured at a distance of 400mm from the sample}
\label{fig:mode}
\end{figure}

\subsection{THz low-pass filter}

\begin{figure}[ht!]
\centering
\includegraphics[width=\linewidth]{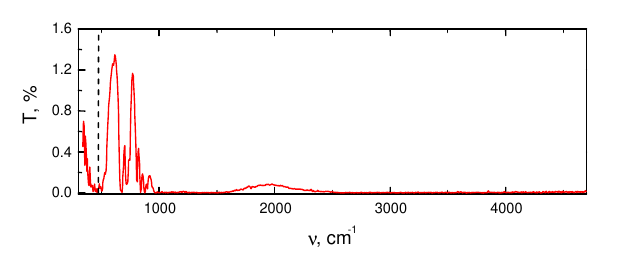}
\caption{Measured IR transmittance of the THz low-pass filter.}
\label{fig:filter}
\end{figure}

In order to reject the shorter wavelentgths that may be present in a 2-color plasma based THz source, we've applied a filter made of layered blackened polyethylene. It's properties were characterized by FTIR spectrometry (SHIMADZU) and the corresponding plot is shown in Fig. S~\ref{fig:filter}. The filter is opaque for wavenumbers down to 450 cm$^{-1}$ below which (marked by the dashed vertical line) there is a sharp onset of transparency. Consequently the filter acts as a low-pass blocking light below approximately 22 um while transmitting THz wavelengths.

\end{document}